\documentclass[usegraphicx,useAMS,usenatbib,]{mn2e}
\usepackage{aas_macros}

\newcommand{\kms}{\mbox{${\rm km\,s}^{-1}$}}
\newcommand{\ace}{\mbox{$\alpha_{\rm CE}$}}
\newcommand{\gmag}{\mbox{$g$}}
\newcommand{\ug}{\mbox{$u-g$}}
\newcommand{\gr}{\mbox{$g-r$}}
\newcommand{\fg}{\mbox{$f-g$}}
\newcommand{\Msolar}{\mbox{${M}_{\sun}$}}
\newcommand{\Rsolar}{\mbox{${R}_{\sun}$}}

\newcommand{\Vsini}{\mbox{${\rm V}_{\rm rot}\sin i$}}

\title[A survey for PCEBs using GALEX and SDSS ]
  {A survey for post common-envelope binary stars using GALEX and
SDSS photometry
\footnote{Based on observations made with the Isaac Newton Telescope
operated on the island of La Palma by the Isaac Newton Group in the Spanish
Observatorio del Roque de los Muchachos of the Instituto de Astrofísica de
Canarias. }}
\author[P.F.L. Maxted et~al.]
{P.F.L.~Maxted$^1$, B.T.~G\"{a}nsicke$^2$, M.R.~Burleigh$^3$, J.~Southworth$^2$,
\newauthor{T.R.~Marsh$^2$, R.~Napiwotzki$^4$, G.~Nelemans$^5$, P.L.~Wood$^1$ }\\
  $^1$Astrophysics Group,  Keele University, Keele, 
      Staffordshire ST5 5BG, United Kingdom\\
 $^2$Department of Physics, University of Warwick, Coventry CV4 7AL\\
 $^3$Department of Physics and Astronomy, University of Leicester, University
Road, Leicester LE1 7RH\\
 $^4$Centre for Astrophysics Research, STRI, University of Hertfordshire,
College Lane, Hatfield AL10 9AB\\
 $^5$Department of Astrophysics, IMAPP, Radboud University Nijmegen, PO Box
9010, 6500 GL Nijmegen, The Netherlands
}
\date{Submitted 2009}

\pagerange{\pageref{firstpage}--\pageref{lastpage}} \pubyear{2008}

\def\LaTeX{L\kern-.36em\raise.3ex\hbox{a}\kern-.15em
    T\kern-.1667em\lower.7ex\hbox{E}\kern-.125emX}

\begin{document}
\label{firstpage}

\maketitle

\begin{abstract}
 We report the first results of our programme to obtain multi-epoch radial
velocity measurements of stars with a strong far-UV excess to identify post
common-envelope binaries (PCEBs).  The targets have been identified using
optical photometry from SDSS DR4, ultraviolet photometry from GALEX GR2 and
proper motion information from SDSS DR5. We have obtained spectra at two or
more epochs for 36 targets. Three of our targets show large radial velocity
shifts ($>50\,\kms$) on a timescale of hours or days and are almost certainly
PCEBs. For one of these targets (SDSS~J104234.77+644205.4) we have obtained
further spectroscopy to confirm that this is a PCEB with an orbital period of
4.74\,h  and semi-amplitude $K =165\,\kms$. Two targets are rapidly rotating
K-dwarfs which appear to show small radial velocity shifts and have strong
Ca\,II H+K emission lines. These may be wind-induced rapidly rotating
(WIRRing) stars. These results show that we can use GALEX and SDSS
photometry to identify PCEBs that cannot be identified using SDSS photometry
alone, and to identify new WIRRing stars.  A more comprehensive survey of
stars identified using the methods developed in this paper will lead to a much
improved understanding of common envelope evolution. 
 
\end{abstract}


\section{Introduction}

 Post common-envelope binary stars (PCEBs) can be loosely defined as short
period binary stars containing a compact object. Cataclysmic variable stars
(CVs), in which a white dwarf accretes matter from a low-mass star that fills
it Roche lobe, are one example of PCEBs. Compact objects are generally the
result of stellar evolution involving at least one phase when the star is a
red giant, but the size of a giant star exceeds the current separation of the
stars in a PCEB by a few orders of magnitude.  \citet{1976IAUS...73...75P}
outlined the solution to this puzzle in a short paper on the origins of
CVs that describes the main features of what is
commonly known as common envelope evolution. In the case of CVs, the evolution
begins with a star of a few solar masses with a less massive companion in a
wide orbit. The evolution of the more massive star through the red giant phase
is terminated prematurely when it exceeds the size of its Roche lobe. The
resulting mass transfer onto the low-mass companion is dynamically unstable,
so the companion is unable to accrete the material from the red giant. The
material forms a common envelope around the dense core of the red giant and
the low-mass companion. If the companion is not massive enough to force the
envelope to co-rotate then the evolution proceeds on a dynamical timescale as
dynamical friction forces the low-mass companion to spiral in through the
outer layers of the red giant. The orbital energy lost by the companion is
transfered to the outer layers of the red giant, which are ejected to reveal
the hot, degenerate core of the red giant and, if it survives, the low-mass
companion. Angular momentum loss by a combination of a magnetic stellar wind
and gravitational wave radiation, or the evolution of the companion may lead
to a second phase of Roche lobe overflow, this time from the low-mass
companion onto the white dwarf, i.e., the formation of a CV. For this reason,
detached binary stars containing a low mass star and white dwarf with an
orbital period of about a day or less are known as pre-CVs
\citep{2003A&A...406..305S}.

The compact object in a PCEB may be a neutron star (NS), black hole (BH) or
white dwarf (WD). Short period binary stars containing hot subdwarf stars (sdO
or sdB stars) are also considered to be PCEBs. The companion star in a PCEB
may be a non-degenerate star or another compact object, e.g., binary pulsars
(NS+NS), ultra-compact X-ray binaries (WD+NS) or AM~CVn binaries (WD+WD).

\begin{figure}
\includegraphics[width=0.46\textwidth]{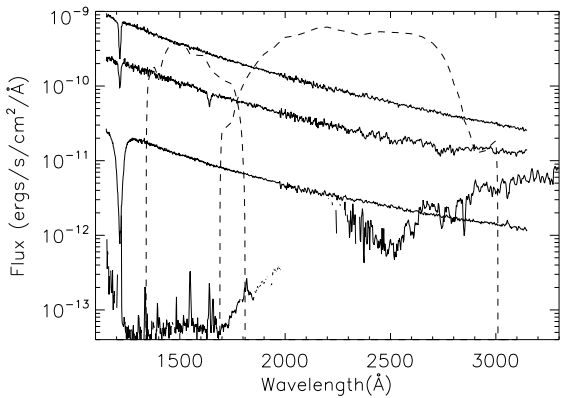}
\caption{The GALEX FUV and NUV response (arbitrary units) compared to IUE
spectra of 3 white dwarfs (HD~149499\,B DOZ1; G191-B2B, DA1 and EG\,274, DA2),
and an active solar-type star EK~Dra (in order of decreasing FUV flux). All
fluxes scaled to a distance of 10\,pc. \label{iueflux} } 
\end{figure}

 There is a vast literature on the theory of the common envelope phase and
many published models of the population of PCEBs \citep{2006LRR.....9....6P,
1993PASP..105.1373I, 2000ARA&A..38..113T}. Considerable debate continues over
fundamental aspects of the CE phase, much of it focused on the definition and
value of the parameter $\ace=\frac{E_{\rm env}}{\Delta E_{\rm orb}}$, where
$\Delta E_{\rm orb}$ is the change in orbital energy of the binary during the
CE phase and $E_{\rm env}$ is the change in the binding energy of the red
giant envelope. In theory,  the outcome of the CE phase for a given binary can
be calculated given a known value for the parameter \ace. In practice, it is
unclear whether the definition of $E_{\rm env}$ should include energy sources
other than the gravitational potential energy of the envelope
\citep{2003MNRAS.343..456S, 2000A&A...360.1043D,  2002MNRAS.336..449H,
2007MNRAS.375.1000B, 2008spbs.conf..233W}, whether evolution prior to Roche
lobe overflow is important \citep{2002ApJ...575.1037E, 2004NewA....9..399S} or
whether a balance of energies gives a valid description of the CE phase
\citep{2000A&A...360.1011N, 2005MNRAS.356..753N}.  These debates are driven by
the discovery and characterization of PCEBs whose properties, either
individually or as a group, are not easily explained by a simple model of the
CE phase, e.g., detached WD+WD ``double degenerate'' binaries
\citep{2002MNRAS.334..833M, 2002MNRAS.332..745M} and sdB stars
\citep{2004Ap&SS.291..307M}. \citet{2009arXiv0903.4152D} have recently
compared the results of binary population synthesis models for a range of
input parameters and using different parametrizations of the common envelope
phase to the observed properties of 35 PCEBs.  They find that the standard
\ace\ parametrization can account for the observed properties of PCEBs with
late-type companions to (pre-)white dwarfs, but cannot explain IK Peg which
has an A8 companion star. They conclude that ``the detection of more PCEBs
with early type secondaries may shed further light on the CE phase''. They
also find that there is a sharp decline in the observed number of PCEBs with
periods $>$ 1 day that cannot be reproduced by any of the models considered,
even if the selection biases against longer period systems are accounted for.
This suggests there are some important features of the common envelope phase
that are not included in the current models.

 One way to better understand the common envelope (CE) phase is to compare the
predictions of binary population synthesis models to the properties of a
large, homogeneously selected sample of PCEBs for which the selection effects
are well understood. This has the advantage of giving a result that is valid
for the general population of that type of PCEB. Comparison of models to one
or two systems may lead to spurious conclusions if one of the systems has had
a peculiar history, e.g., V471~Tau appears to be a normal PCEB containing a
white dwarf and a K-dwarf, but its membership of the Hyades enables us to
infer that it has a peculiar evolutionary history and is probably the result
of the evolution of a triple star system  \citep{2001ApJ...563..971O}.

 The Sloan Digital Sky Survey (SDSS) makes it possible to construct  a
well-defined sample of PCEBs that can be identified from their optical
colours. This is particularly so for non-interacting binaries containing a WD
with M-dwarf companions (WD+M).  These binaries have distinctive colours that
enable them to be efficiently identified from the SDSS photometry alone,
\citep{2008A&A...486..843A, 2007ASPC..372..459S, 2009AJ....137.4377Y} and
there is also spectroscopy of many such systems from the same survey
\citep{2008arXiv0811.1508H, 2007MNRAS.382.1377R}. There are also many CVs with
SDSS spectra that have been targeted for observation from their SDSS colours
or observed serendipitously \citep{2007AJ....134..185S}.  The selection
effects for CVs are large and depend strongly on their accretion properties,
whose dependence on the other properties of the binary are themselves poorly
understood. For this reason, detached PCEBs are a better choice than CVs as a
sample for testing models of CE evolution for WD binaries.

 Using SDSS photometry alone it is only possible to detect unresolved
non-interacting binaries if the companion to the white dwarf has a  spectral
type later than about M0, the exact limit depending on the temperature of the
white dwarf \citep{2008A&A...486..843A}. \citet {2003A&A...406..305S}
calculated the temperature of the coolest WD that can be detected from a
U-band excess as function of spectral type for a main-sequence companion. The
limits are approximately 45,000K, 28,000K and 9,000K for spectral types K5, M0
and M6, respectively. The limits will be similar for the SDSS or any survey
based on optical photometry because for cooler WDs the main-sequence companion
is much brighter than a non-accreting WD at all optical wavelengths. 

\begin{figure*}
\includegraphics[width=0.96\textwidth]{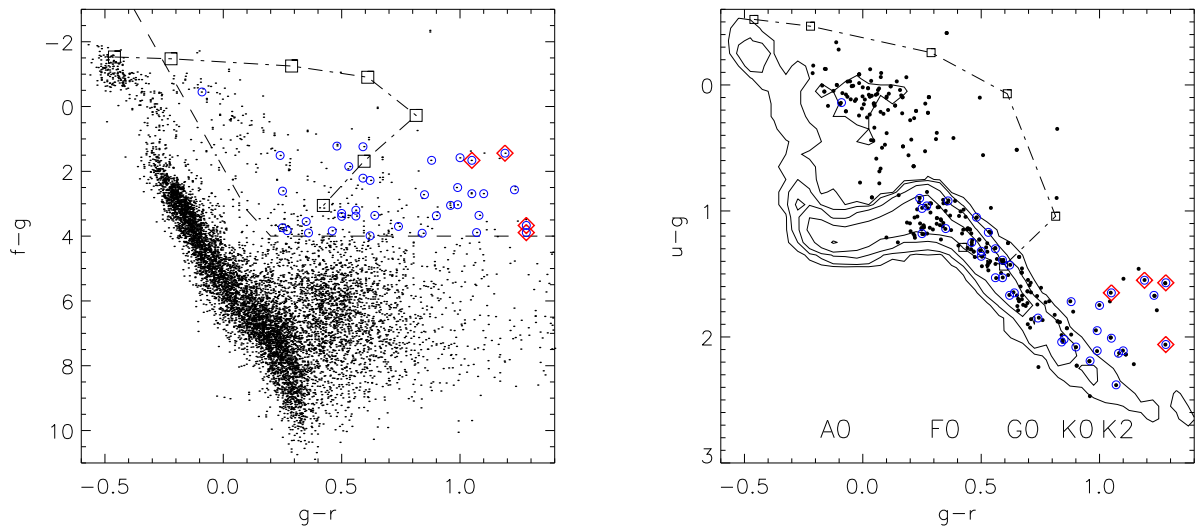}
\caption{Colour-colour diagrams for our sample of stars with reliable optical
and ultraviolet photometry. The limit for selecting stars with an ultraviolet
excess is shown in the left panel as a dashed line. Stars above this line are
plotted with dots in the right panel.  For clarity, we show the density of
points in the right panel with contours rather than individual points. An
approximate indication of spectral type for single stars based on their $g-r$
colour is indicated in the bottom of the right panel. Stars for which we have
obtained spectra are circled and stars with variable radial velocities are
marked with diamonds. The dash-dot line with squares shows the combined
colours of a white dwarf with T$_{\rm eff}=$48,000K  plus cool companions with
spectral types from G2 to M5 (the blue-most point is the white dwarf alone).
\label{colcol}} \end{figure*}

 To find non-accreting white dwarfs with F-, G- and K-type companions or to
find cool WDs with M-type companions requires a survey at UV or soft X-ray
wavelengths. \citet{1996PASP..108..250L} identified two late-type stars with
hidden hot white dwarfs (56~Per and HR~3643) from follow-up observations of
late-type stars with a 1565\AA\ excess in the TD-1 ultraviolet sky survey.
These are Sirius-like binaries, i.e., non-interacting systems with large
orbital separations. Many more Sirius-like binaries were identified as a
result of the EUVE and ROSAT surveys \citep{2002ASPC..264...27B}. These surveys
also identified several new PCEBs \citep{1997MNRAS.287..381B,
1994MNRAS.270..499B, 1998ApJ...502..763V}. There have also been serendipitous
discoveries, e.g.,  from IUE spectroscopy \citep{1993AJ....106.1113B,
1992AJ....104.1539B}. These serendipitous discoveries can provide useful
evidence for the existence of a particular type of binary star, but it is
difficult to know whether they should be included in any test of a binary
population synthesis model.

 The advent of the GALEX all-sky UV survey \citep{2005ApJ...619L...1M} makes
it feasible to conduct a large survey to detect hot white dwarfs in unresolved
binaries with G-type and K-type companions and cool WD with early-M type
companions.  In this paper we report the first results of our survey to obtain
follow-up spectroscopy of solar-type stars with a far-UV excess identified using
GALEX and SDSS photometry.  The main aim of the survey is to produce a
well-defined sample of detached PCEBs that cannot be identified using SDSS
photometry alone. The properties of these PCEBs will be a strong test of
population synthesis models.  We also expect that this survey will find other
interesting objects, e.g., CVs and other interacting binary stars.

\section{Target selection}

 To discover PCEBs in which the non-degenerate companion dominates the optical
flux we require a sample of stars with reliable optical and ultraviolet
photometry. We also require that the sources are bright so that the follow-up
of the targets can be done efficiently. To produce this sample of stars we
used optical SDSS photometry and ultraviolet GALEX photometry. We started with
the table provided by Multi-mission Archive at STScI of objects from SDSS DR4
cross matched within 4\,arcsec of objects in GALEX GR2 (table  xSdssDr4).
From this table we selected objects satisfying the following criteria:
\begin{itemize}
\item  $<0.6^\circ$ from the centre of the GALEX field of view;
\item the closest SDSS object to the GALEX source position;
\item GALEX magnitudes $n<25$ and $f<25$;
\item SDSS $g$-magnitude $g<16.5$;
\item SDSS morphological type class 6 (STAR);
\item ``clean'' SDSS photometry;\footnote{
{\it See http://skyserver.sdss.org/dr2/en/help/docs/\\~realquery.asp\#flags}}
\item proper motion  $>0.01$\,arcsec/yr.
\end{itemize}
The $g$ magnitudes used in this paper are taken from the SDSS table entry
psfMag\_g, and similarly for other SDSS magnitudes. The resulting table has
2960 sources. Although the matching radius is 4\,arcsec, the positions for GR2
and DR4 agree to within 1\,arcsec for 80\% of the sources.  The restriction on
the source position in the field of view is to avoid spurious detections that
are common in these regions \citep{ 2005AJ....130.1022A}. The proper motion
information was taken from a table provided by SDSS DR5 and is based on the
astrometry from the SDSS plus recalibrated USNO-B astrometry. Objects detected
at fewer than 6 epochs in SDSS plus USNO-B were excluded. This restriction on
the proper motion was included to reduce the contamination of the sample by
background sources (galaxies and quasars). Without this criterion the sample
size would be 8760 sources. The disadvantage of adding this proper motion
criterion is that it introduces a kinematical bias into the survey. A complete
treatment of this bias is beyond the scope of this paper.

 The majority of objects in this sample of 2960 sources are expected to be
moderately bright stars with reliable optical and ultraviolet photometry.
There are no stars brighter than $g = 13$ in the sample because  stars brighter
than this appear saturated in the SDSS images so they do not have reliable
SDSS photometry. 

 The far-UV (FUV) and near-UV (NUV) passbands of the GALEX instrument are
shown in Fig.~\ref{iueflux} and are compared to the UV spectra of three WDs and
a very active G0V star. It can be seen that hot WDs (DA2 or hotter) can be
easily distinguished from an active solar-type stars at FUV wavelengths.  

NUV  and $u$-band fluxes can be strongly affected by chromospheric activity in
late-type stars (Fig.~\ref{iueflux}), so  we used the $f-g$ colour to identify
stars with a UV excess.  The distribution of these sources in the \fg\ v. \gr\
and \ug\ v. \gr\  colour-colour diagrams are shown in Fig.~\ref{colcol}. There
is a well defined main-sequence relation in the optical colour-colour diagram,
showing that the optical photometry for the majority of stars in our sample is
reliable. Our criteria for selecting stars with a UV excess as shown in this
figure are \[ (f-g) < 13(g-r)+1.9 {\rm ~~and~~} (f-g) < 4. \] There are 90
targets that satisfy these selection criteria.  

\subsection{Cross-identifications for target objects}
 We have used the SIMBAD database to search for catalogued objects matching
our target positions.  The results are given in Table~\ref{simbad} and some
individual objects are discussed below. These stars give some impression of
the type of object that may be discovered using our survey. In most cases, but
not all (e.g. U~Sex), we avoided observing objects where the nature of the
star was already clear from existing observations. 

\begin{description}
\item [\bf KUV 03134$-$0001]{The spectral type of this star given by
\citet{1993AJ....106..390W} is
``cont. + dM -- Composite spectrum; shows blue continuum plus TiO features in
the red''. }
\item [\bf PQ~Gem]{A well-known intermediate polar \citep{2006MNRAS.369.1229E,
1993MNRAS.264..171R}.}
\item [\bf J0852+0313]{The SDSS spectrum of this object shows that it is a
QSO with a redshift $z=0.297$. There is a faint companion within a few
arcseconds to this QSO visible in the SDSS images that is unresolved in the
DSS images from which the USNO-B positions are measured. We conclude that the
proper motion value for this object is spurious.}
\item [\bf BZ UMa]{This is a dwarf nova \citep[CV,][]{2006MNRAS.369..369N}.}
\item [\bf PG 1056+324]{\citet{1988ApJ...328..213W} give a spectral type of
``sdB'' for this star.}
\item [\bf BE UMa]{ This is a well-studied  detached PCEB containing a very
hot (pre-)white dwarf and a low-mass, non-degenerate companion with an orbital
period of 2.3\,d \citep{1999ApJ...518..866F}.}
\item [\bf SDSS J140916.11+382832.1]{The automatic classification algorithm
of \citet{2006ApJS..167...40E} identifies this star  as a hot subdwarf. Visual
inspection of the SDSS spectrum shows  hints of absorption lines due to a
late-type companion (G-band, Mg-b, Ca\,II IR triplet). }
\item [\bf SBSS 1422+497]{The spectral type given by \citet{2005RMxAA..41..155S}
 is ``DA''.}
\item [\bf PG1657+416]{This is a pulsating sdB star with a G5 main-sequence
companion star \citep{2007A&A...461..585O}.}
\end{description}

\begin{figure*}
\includegraphics[width=0.98\textwidth]{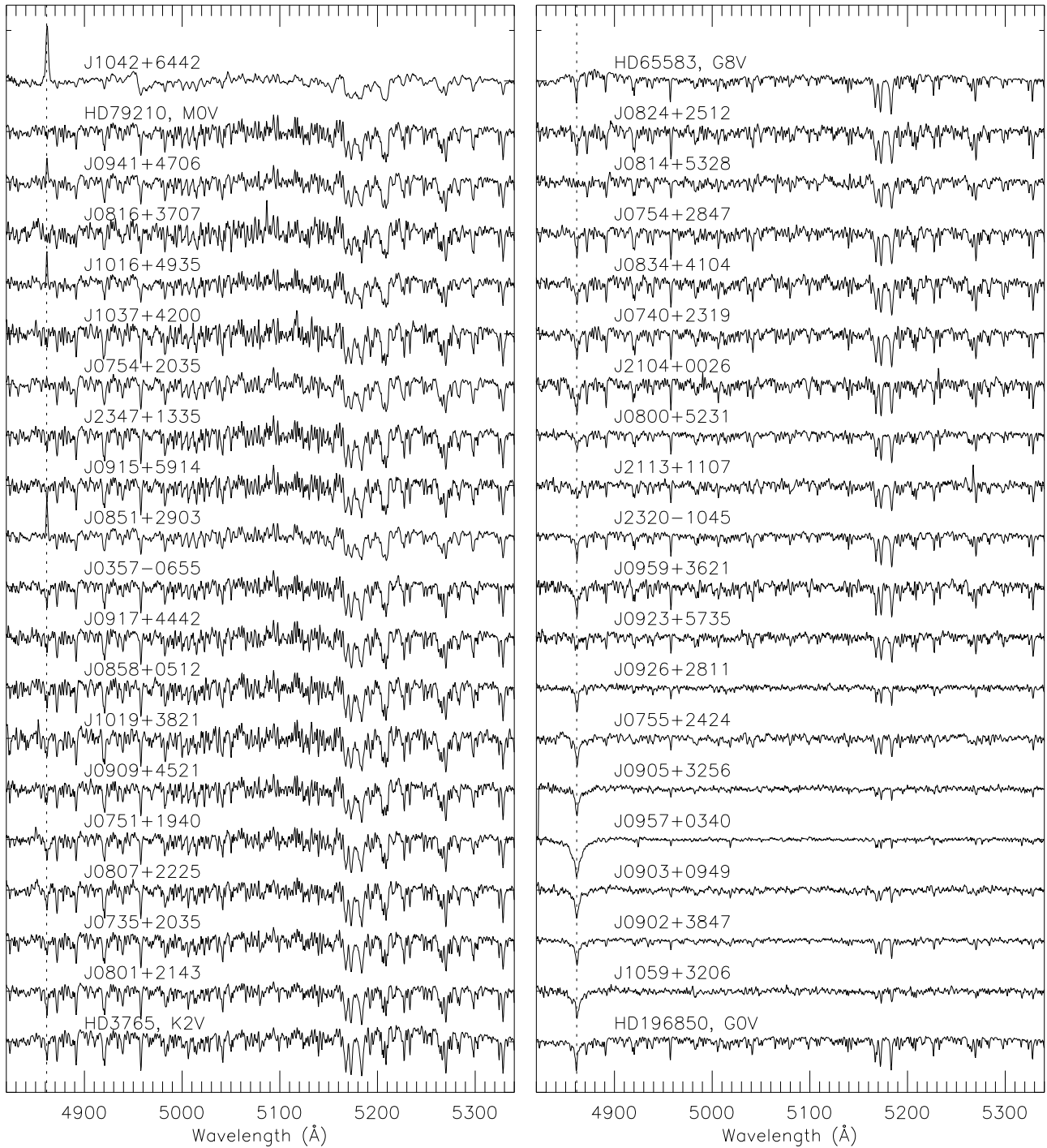}
\caption{Spectra of our targets ordered by their $g-r$ colour. Spectra have
been co-added on a common wavelength scale, normalized, smoothed and re-binned
for display and are offset by 1 for clarity. Spectra of some radial velocity
standard stars with known spectral types as shown in the label are included
for comparison. The rest wavelength of the H$\beta$ line is indicated with a
dotted line. \label{spectra}} 
\end{figure*}

\begin{table*}
\begin{minipage}{0.95\textwidth}
\caption{Cross-identifications for objects from our sample with sources listed by
SIMBAD. \label{simbad}}
\begin{tabular}{@{}lrrrrrrll}
\hline
ID & \multicolumn{1}{c}{$\alpha_{2000}$}
&\multicolumn{1}{c}{$\delta_{2000}$}
&\multicolumn{1}{c}{\gmag} & \multicolumn{1}{c}{\ug}
&\multicolumn{1}{c}{\gr} & \multicolumn{1}{c}{\fg} & Other ID & Notes\\
\hline
J0316+0009 & 03 16 00.2 & +00 09 47 & 16.45 &$ 0.52 $&$ 0.65 $&$ 0.06 $& KUV 03134-0001 \\
J0751+1444 & 07 51 17.3 & +14 44 24 & 14.18 &$-0.06 $&$ 0.18 $&$ 1.44 $& PQ
Gem & Intermediate polar \\
J0824+3028 & 08 24 34.1 & +30 28 55 & 15.25 &$ 0.09 $&$-0.07 $&$-0.44 $& PG
0821+306 & QSO \\
J0852+0313 & 08 52 59.2 & +03 13 21 & 16.22 &$ 0.08 $&$ 0.06 $&$ 1.30 $& 2MASS J08525922+0313207  \\
J0853+5748 & 08 53 44.2 & +57 48 41 & 16.40 &$-0.41 $&$ 0.35 $&$ 0.74 $& BZ
UMa & Dwarf nova \\
J0941+4706 & 09 41 55.2 & +47 06 38 & 16.12 &$ 2.06 $&$ 1.28 $&$ 3.89 $& XMS J094154.6+470637  \\
J0957+0340 & 09 57 25.4 & +03 40 05 & 14.91 &$ 1.18 $&$ 0.25 $&$ 2.61 $& U Sex
& RR Lyrae star \\
J1014+0033 & 10 14 54.9 & +00 33 37 & 16.27 &$ 0.08 $&$ 0.16 $&$ 0.66 $& QSO B1012+008  \\
J1042+6442 & 10 42 34.8 & +64 42 05 & 15.00 &$ 1.57 $&$ 1.28 $&$ 3.67 $& 1RXS J104237.0+644217  \\
J1054+3600 & 10 54 50.6 & +36 00 13 & 16.48 &$ 2.47 $&$ 0.96 $&$ 3.12 $& LEDA 3088408   \\
J1059+3206 & 10 59 05.2 & +32 06 21 & 14.64 &$ 0.14 $&$-0.09 $&$-0.45 $& PG
1056+324 & sdB \\
J1121+0012 & 11 21 19.7 & +00 12 12 & 14.90 &$ 1.18 $&$ 0.42 $&$ 3.38 $& 2MASS J11211965+0012120  \\
J1155+0640 & 11 55 28.1 & +06 40 18 & 15.70 &$ 1.05 $&$ 0.43 $&$ 1.91 $& Pul$-$3 850246  \\
J1157+4856 & 11 57 44.8 & +48 56 18 & 15.78 &$-0.28 $&$-0.10 $&$-1.72 $& BE
UMa  & PCEB, P=2.3\,d\\
J1300+4139 & 13 00 40.1 & +41 39 37 & 15.07 &$ 0.94 $&$ 0.34 $&$ 3.05 $& BPS BS 16938-0017  \\
J1409+3828 & 14 09 16.1 & +38 28 32 & 16.03 &$-0.13 $&$-0.16 $&$-1.03 $& SDSS J140916.11+382832.1 \\
J1418+3705 & 14 18 31.8 & +37 05 41 & 15.41 &$ 1.27 $&$ 0.46 $&$ 2.99 $& 2MASS J14183182+3705411  \\
J1424+4929 & 14 24 40.5 & +49 29 58 & 16.24 &$ 0.10 $&$ 0.49 $&$-1.06 $& SBSS
1422+497  & DA\\
J1658+4131 & 16 58 41.8 & +41 31 16 & 15.93 &$-0.09 $&$-0.21 $&$-0.86 $& PG
1657+416 & sdB+G5V\\
\hline
\end{tabular}
\end{minipage}
\end{table*} 

\begin{table*}
\begin{minipage}{0.95\textwidth}
\caption{Summary of our radial velocity measurements. N$_{\rm rv}$ is the
number of radial velocity measurements. $\Delta t$ is the baseline of the
observations. See text for the definition of $\log p$. FWHM is the mean
full-width at half-maximum of the CCF. Where a $\star$ symbol appears in the
final column further remarks on the star appear in Section~\ref{Notes}.
\label{RVsummary}}
\begin{tabular}{@{}lrrrrrrrrrrrl}
\hline
ID & \multicolumn{1}{c}{RA}
&\multicolumn{1}{c}{Dec}
&\multicolumn{1}{c}{\gmag} & \multicolumn{1}{c}{\ug}
&\multicolumn{1}{c}{\gr} & \multicolumn{1}{c}{\fg}
&\multicolumn{1}{c}{N$_{\rm rv}$} & \multicolumn{1}{c}{$\log p$}
& \multicolumn{1}{c}{$\Delta t$}&\multicolumn{1}{c}{EW(H$\beta$)} 
& \multicolumn{1}{c}{FWHM} & \\
& (J2000)& (J2000)& & & & & & & (days) &(\AA) & (\kms) & \\
\hline
J0357$-$0655 & 03:57:16.0 &$-$06:55:04 & 16.07 & 1.75 &$ 1.00 $&$ 1.58 $& 5&$ -0.10 $ & 693.1 &$  1.4 $&  70& \\
J0735$+$2035 & 07:35:34.5 &$+$20:35:32 & 15.87 & 2.04 &$ 0.84 $&$ 3.91 $& 5&$ -0.51 $ & 696.1 &$  0.7 $&  67& \\
J0740$+$2319 & 07:40:33.6 &$+$23:19:26 & 14.56 & 1.53 &$ 0.59 $&$ 2.21 $& 2&$ -0.02 $ & 691.1 &$  1.8 $&  65& \\
J0751$+$1940 & 07:51:32.0 &$+$19:40:55 & 16.46 & 1.72 &$ 0.88 $&$ 1.66 $& 4&$  0.00 $ & 691.2 &$  1.5 $&  69& \\
J0754$+$2035 & 07:54:07.5 &$+$20:35:52 & 16.12 & 2.13 &$ 1.08 $&$ 3.36 $& 6&$ -1.12 $ & 696.1 &$ -0.4 $& 113& $\star$ \\
J0754$+$2847 & 07:54:04.2 &$+$28:47:01 & 16.10 & 1.67 &$ 0.62 $&$ 3.99 $& 2&$ -0.25 $ & 691.2 &$  1.4 $&  67& \\
J0755$+$2424 & 07:55:41.9 &$+$24:24:36 & 16.27 & 1.14 &$ 0.35 $&$ 3.55 $& 2&$ -0.35 $ & 692.1 &$  2.4 $&  86& \\
J0800$+$5231 & 08:00:38.6 &$+$52:31:29 & 16.28 & 1.30 &$ 0.56 $&$ 3.20 $& 4&$ -2.69 $ & 696.0 &$  1.1 $&  69& $\star$ \\
J0801$+$2143 & 08:01:07.1 &$+$21:43:37 & 15.09 & 1.85 &$ 0.74 $&$ 3.70 $& 2&$ -0.17 $ & 691.2 &$  1.2 $&  69& \\
J0807$+$2225 & 08:07:36.5 &$+$22:25:37 & 16.21 & 2.02 &$ 0.85 $&$ 2.72 $& 2&$ -0.03 $ & 691.2 &$  1.0 $&  73& \\
J0814$+$5328 & 08:14:43.4 &$+$53:28:05 & 16.48 & 1.43 &$ 0.62 $&$ 2.28 $& 2&$ -0.23 $ & 692.0 &$  0.1 $&  92& $\star$ \\
J0816$+$3707 & 08:16:07.1 &$+$37:07:04 & 16.39 & 1.67 &$ 1.23 $&$ 2.57 $& 2&$ -0.11 $ & 691.1 &$  0.9 $&  69& \\
J0824$+$2512 & 08:24:56.5 &$+$25:12:47 & 15.83 & 1.65 &$ 0.64 $&$ 3.36 $& 2&$ -0.25 $ & 691.2 &$  1.7 $&  67& \\
J0834$+$4104 & 08:34:50.0 &$+$41:04:23 & 15.73 & 1.39 &$ 0.59 $&$ 1.24 $& 2&$ -0.14 $ & 691.2 &$  1.5 $&  65& \\
J0851$+$2903 & 08:51:37.2 &$+$29:03:30 & 16.24 & 1.65 &$ 1.05 $&$ 1.66 $& 9&$ -2.16 $ & 696.1 &$ -1.3 $& 120& \\
J0858$+$0512 & 08:58:03.2 &$+$05:12:49 & 15.84 & 2.11 &$ 0.99 $&$ 3.03 $& 2&$ -0.11 $ & 691.3 &$  1.1 $&  69& \\
J0902$+$3847 & 09:02:04.1 &$+$38:47:38 & 14.74 & 0.90 &$ 0.24 $&$ 1.51 $& 6&$ -0.99 $ & 693.2 &$  2.1 $&  76& \\
J0903$+$0949 & 09:03:51.2 &$+$09:49:35 & 16.38 & 0.98 &$ 0.25 $&$ 3.74 $& 3&$ -0.09 $ & 4.1   &$  2.7 $& 121& \\
J0905$+$3256 & 09:05:24.4 &$+$32:56:02 & 15.24 & 0.96 &$ 0.27 $&$ 3.83 $& 4&$ -0.37 $ & 4.0   &$  2.1 $&  76& $\star$ \\
J0909$+$4521 & 09:09:38.0 &$+$45:21:42 & 15.63 & 2.08 &$ 0.90 $&$ 3.37 $& 3&$ -0.21 $ & 4.1   &$  0.5 $&  71& \\
J0915$+$5914 & 09:15:42.7 &$+$59:14:57 & 15.84 & 2.01 &$ 1.05 $&$ 2.69 $& 3&$ -0.09 $ & 4.0   &$  0.7 $&  74& \\
J0917$+$4442 & 09:17:10.3 &$+$44:42:26 & 15.84 & 1.95 &$ 0.99 $&$ 2.50 $& 4&$ -0.07 $ & 4.0   &$  0.6 $&  69& \\
J0923$+$5735 & 09:23:23.1 &$+$57:35:40 & 15.25 & 1.05 &$ 0.48 $&$ 1.22 $& 3&$ -0.12 $ & 4.1   &$  0.4 $&  60& \\
J0926$+$2811 & 09:26:16.6 &$+$28:11:13 & 15.98 & 0.92 &$ 0.36 $&$ 3.90 $& 4&$ -0.03 $ & 4.1   &$  1.6 $&  61& \\
J0941$+$4706 & 09:41:54.6 &$+$47:06:38 & 16.12 & 2.06 &$ 1.28 $&$ 3.89 $& 4&$ < -40 $ & 4.0   &$ -1.5 $&  96& $\star$ \\
J0957$+$0340 & 09:57:25.4 &$+$03:40:05 & 14.91 & 1.18 &$ 0.25 $&$ 2.61 $& 3&$ -0.36 $ & 1.1   &$  4.3 $&  76& \\
J0959$+$3621 & 09:59:02.4 &$+$36:21:03 & 14.49 & 1.36 &$ 0.50 $&$ 3.39 $& 2&$ -0.07 $ & 1.0   &$  2.3 $&  58& \\
J1016$+$4935 & 10:16:57.1 &$+$49:35:14 & 16.11 & 1.55 &$ 1.19 $&$ 1.44 $& 4&$ < -40 $ & 3.9   &$ -0.8 $&  71& $\star$ \\
J1019$+$3821 & 10:19:37.4 &$+$38:21:12 & 15.41 & 2.19 &$ 0.96 $&$ 3.05 $& 2&$ -0.46 $ & 1.0   &$  0.9 $&  68& \\
J1037$+$4200 & 10:37:08.3 &$+$42:00:39 & 15.78 & 2.11 &$ 1.10 $&$ 2.69 $& 2&$ -0.07 $ & 1.0   &$  0.3 $&  69& \\
J1042$+$6442 & 10:42:34.8 &$+$64:42:05 & 15.00 & 1.57 &$ 1.28 $&$ 3.67 $& 9&$ < -40 $ & 3.0   &$ -5.1 $& 180& $\star$ \\
J1059$+$3206 & 10:59:05.2 &$+$32:06:21 & 14.64 & 0.14 &$-0.09 $&$-0.45 $& 2&$ -1.06 $ & 2.9   &$  2.5 $&  66& \\
J2104$+$0026 & 21:04:01.4 &$+$00:26:54 & 16.29 & 1.53 &$ 0.56 $&$ 3.38 $& 2&$ -0.05 $ & 4.0   &$  2.7 $&  57& \\
J2113$+$1107 & 21:13:16.2 &$+$11:07:48 & 16.31 & 1.17 &$ 0.53 $&$ 1.85 $& 2&$ -0.24 $ & 4.0   &$  1.5 $&  70& \\
J2320$-$1045 & 23:20:40.4 &$-$10:45:11 & 15.22 & 1.32 &$ 0.50 $&$ 3.30 $& 6&$ -0.02 $ & 5.0   &$  1.6 $&  63& \\
J2347$+$1335 & 23:47:42.4 &$+$13:35:57 & 15.54 & 2.38 &$ 1.07 $&$ 3.89 $& 4&$ -0.01 $ & 5.0   &$  0.6 $&  69& \\
\noalign{\smallskip}
\hline
\noalign{\smallskip}
\end{tabular}
\end{minipage}
\end{table*} 

\subsection{Observations}
 We obtained spectroscopy for 36 targets from our sample using the
Intermediate Dispersion Spectrograph (IDS) on the 2.5m Isaac Newton Telescope.
We used the H2400B grating with a 1.2\,arcsec slit and the EEV10 charge
coupled device (CCD) detector to obtain spectra with a resolution of 0.45\AA\
sampled at 0.22\AA/pixel. The unvignetted portion of the CCD covers the
wavelength range 4875\,--\,5375\AA. We also extracted the region of the
spectrum around the H$\beta$ line that lies in the vignetted region of the CCD.
Spectra were extracted using the optimal extraction algorithm of
\citet{1986PASP...98..609H}. Observations of each star were bracketed 
with arc spectra and the wavelength calibration established from these arcs
interpolated to the time of mid-exposure. We obtained a single spectrum of 19
of our targets on 2007 Jan 29. The majority of the spectra were obtained
in the interval 2008 Dec 20 to 2008 Dec 26. The average spectrum of each
target is shown in Fig.~\ref{spectra}.

 We measured the radial velocity (RV) of our targets using cross correlation
against a spectrum of the RV standard star HD\,65583 \citep[G8V,
$V_r=+14.85\,\kms$;][]{1999ASPC..185..367U}. The spectral range used for the
cross-correlation was 4868\,--\,5395\AA.  We obtained spectra of the RV standard
star HD\,3765 on every night of the 2007~Jan and 2008~Dec runs to monitor the
stability of our RV measurements. The standard deviation of the RV
measurements from these spectra is 2.79\,\kms. This is much larger than the
precision with which the peak of the cross-correlation function (CCF) can be
located and is likely to be dominated by the motion of the stellar image
within the spectrograph slit. We added this estimate of the external error in
quadrature to the internal error of the RV measurements of our targets to
obtain a total standard error estimate for each RV measurement.   The exposure
times used for our targets (300\,--\,1800s) are much longer than those used
for  HD\,3765 (30\,--\,60s). Guiding errors during a longer exposure will
smear the light of the star across the slit so the external error due to image
motion in the slit is likely to be less for our targets than for HD\,3765,
i.e., the total standard errors we have adopted are likely to be pessimistic. 

 For the set of RV measurements of each target $V_{r,i};~~ i=1,\dots,N_{\rm
RV}$, we calculate the weighted mean RV value, $\overline{V}_r$. To test  for
variability of the RV we calculate the probability, $p$, of obtaining the
observed value of $\chi^2$ or greater for the model $V_{r,i} =
\overline{V}_r$ assuming normally distributed errors, i.e., a small value of $p$
indicates that the RV measurements are not constant. The value of $p$ for each
target are also given in Table~\ref{RVsummary} as $\log p $. 

 Also shown in Table~\ref{RVsummary} is a crude estimate of the equivalent
width of the H$\beta$ line, EW(H$\beta$) in the average (median) spectrum of
the star after shifting the individual spectra to a common wavelength scale
according to their measured RVs. The equivalent width was calculated by
numerical integration in a window $\pm 5$\AA\ wide around the rest wavelength
of the line. Negative values of EW(H$\beta$) indicate that the
H$\beta$ line is in emission.

\section{Results}

\subsection{Detection efficiency}
\begin{figure}
\includegraphics[width=0.46\textwidth]{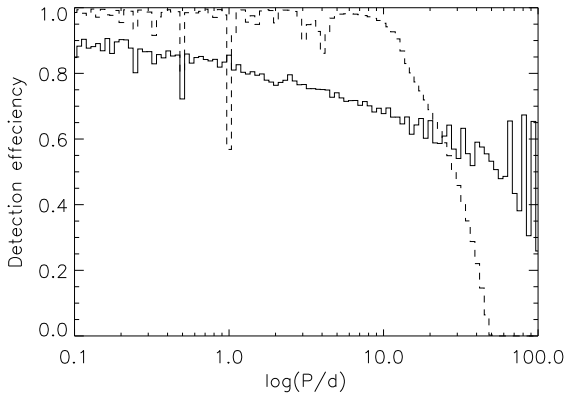}
\caption{The fraction of binaries detected with $\log p  < -3$ assuming a mass
ratio $q=0.75$ and a mass for the visible star of 0.75\Msolar\ using data with
the same sampling and standard errors as that obtained for J0801+2143 (solid
line) and J0909+4521 (dashed line). The detection efficiency in each bin is
the average of 100 periods evenly distributed between the bin limits.
\label{efac}}
\end{figure}

 For each star we calculate the probability of detecting a companion from the
radial velocity variations using the method of \citet{2001MNRAS.326.1391M}. We
assume a mass of 0.75\Msolar\ for the visible star because this is likely to
be close to the average mass of our target stars. We assume a mass ratio of
$q=0.75$ because this value then gives a white dwarf companion mass of
0.56\Msolar, which is a typical mass for a white dwarf. The results for two
stars are shown in Fig.~\ref{efac}. We find that the average detection
efficiency near a period of 10\,d   is 79\% and that the detection efficiency
is more than 50\% for periods of less than 10\,d for all 36 of the stars
observed.

\subsection{Notes on individual objects\label{Notes}}

\begin{figure}
\includegraphics[angle=270,width=0.46\textwidth]{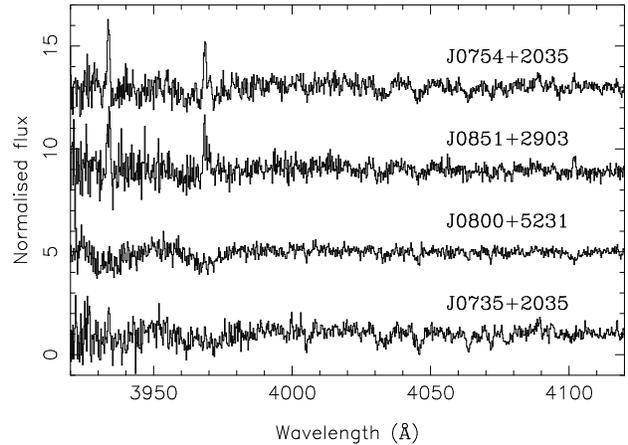}
\caption{Spectra of selected targets in the region of the Ca\,II H and K
lines. The spectra have been normalized, smoothed and vertically offset for
clarity. \label{HK}} 
\end{figure}

\begin{description}
\item[{\bf J0754+2035}]{We obtained a single spectrum of this star with the
IDS and the H2400B grating at a central wavelength of 4200\AA. This spectrum
shows strong emission in the cores of the Ca\,II H and K lines (EW $\approx
-3.5$\AA) for this star (Fig.~\ref{HK}). The width of the CCF in this star
shows this star is a rapid rotator and so is expected to be chromospherically
active. There is some weak evidence for a small amount of RV variability in
this star ($\approx 10\kms$), which may be a result of star spots rather than
binarity. }
\item[{\bf J0800+5231}]{There is an offset of about 10\kms\ between the RVs
measured for this star in 2007 and 2008, but no significant variability over 4
nights from the 2008 data alone. This may be a binary with a long orbital
period. A single spectrum near 4200\AA\ obtained with the IDS (Fig.~\ref{HK})
suggests that the spectral type is late-G, which is consistent with the
$u-g$ and $g-r$ colours of the star.}
\item[{\bf J0814+5328}]{The spectra near the H$\beta$ line are noisy  but the
absorption line does appear to be shallow compared to other stars of the same
colour/spectral-type (Fig.~\ref{spectra}). The absorption line may be
``filled-in'' by weak emission at this wavelength.}
\item[{\bf J0851+2903}]{This is a similar case to J0754+2035. A single
spectrum around 4200\AA\ shows Ca\,II H and K lines in emission
 (EW $\approx -3$\AA, Fig.~\ref{HK}), and there is some evidence of RV
variability at the level of 20\kms\ which may be the result of star spots in
this rapidly rotating star. }
\item[{\bf J0905+3256}]{This star has a large heliocentric RV ($\approx
-320\kms$) so is likely to be a halo star. } 
\item[{\bf J0941+4706}]{This star is an X-ray source detected by the XMS survey
\citep{2007A&A...476.1191B}. The RV changes by more than 50\kms\ in less than
3 hours. This is too large to be explained by star spots. The H$\beta$
emission line has a similar width to the CCF and shows the same RV variation,
i.e., the H$\beta$ emission is due to irradiation of and/or chromospheric
activity on the M-dwarf and is not due to an accretion disc. This is almost
certainly a PCEB. } 
\item[{\bf J0957+0340}]{This is an RR~Lyr-type variable star of type RRab 
\citep{1971GCVS3.C......0K}. The amplitude of the variability of these stars
at UV wavelengths is very large \citep{2005ApJ...619L.123W}, so it is not
surprising that optical and UV photometry obtained at different epochs result
in a peculiar position for the star in the UV-optical colour-colour diagram.}
\item[{\bf J1016+4935}]{The RV of this star changes by  $ \approx 100\kms$
over 3\,d.  The H$\beta$ emission line has a similar width to the CCF and
shows the same RV variation, i.e., the H$\beta$ emission is due to irradiation
of and/or chromospheric activity on the M-dwarf. This is almost certainly a
PCEB. We have too few data to estimate the orbital period. }
\item[{\bf J1042+6442}]{This star is an X-ray source detected by the ROSAT-FSC
survey \citep{2006A&A...449..425M}.  The RV changes by approximately 250\kms\
 in less than 3 hours. Additional spectroscopy has been obtained and is
discussed in the next section.}
\end{description}

\begin{figure}
\includegraphics[angle=270,width=0.46\textwidth]{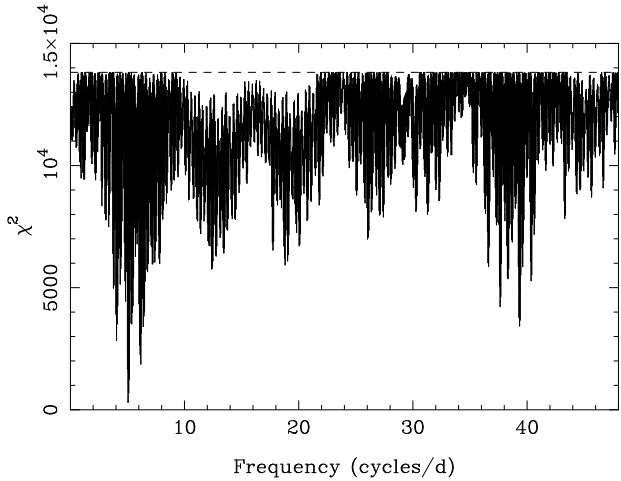}
\caption{Periodogram of our RV measurements for J1042+6442. The value of
$\chi^2$ is shown for a least-squares fit of a sine wave at each
frequency.\label{pgram}}
\end{figure}

\begin{figure}
\includegraphics[width=0.46\textwidth]{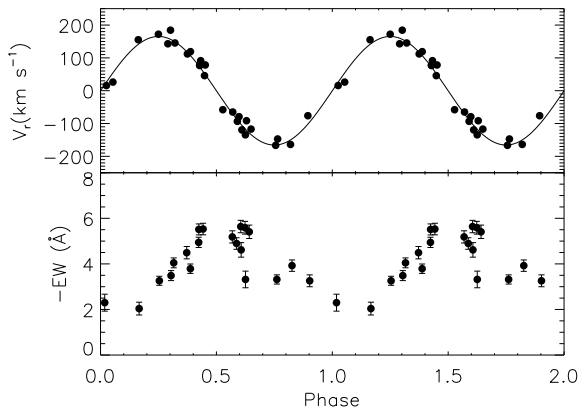}
\caption{{\it Upper panel: } RV measurements of J1042+6442 with the circular
orbit fit by least-squares plotted as a function of orbital phase
for $P=0.19767$\,d. {\it Lower panel: } Equivalent width of the
H$\beta$ emission line. \label{J1042Fig}. Measurements with large uncertainties
have been excluded for clarity. } 
\end{figure}

\subsection{J1042+6442\label{J1042Sec}}
 The data for this star obtained during 2008~Dec clearly show that the RVs 
vary with a period $\sim 0.2\,$d. We obtained further spectroscopy with the
same instrument and configuration during Jan~2009 to measure the orbital
period and mass function of this star. A further 17 spectra were obtained,
although some of these have poor signal-to-noise due to poor seeing. The RV of
J1042+6442 was measured as above for all the spectra and are given in
Table~\ref{J1042RV}.  We then searched for periodicity in these RV
measurements by fitting sine waves at a range of trial frequencies. The period
is found to be close to 0.19767\,d and is unambiguous (Fig.~\ref{pgram}). Also
given in Table~\ref{J1042RV} is the equivalent width of the H$\beta$ emission
line measured by numerical integration in a window $\pm 250\kms$ around the
position of the line based on the measured RV. We used a least-squares fit of
a circular orbit to the RV measurements to derive the parameters shown in
Table~\ref{J1042Orbit}. The scatter of the  RV measurements around the best
fit is larger than expected given the standard errors of the RV measurements.
We account for this by adding a quantity $\sigma_{\rm ex}$ in quadrature to
the standard errors of the RV measurements prior to calculating the fit. The
physical origin of $\sigma_{\rm ex}$ may be related to the extreme
chromospheric activity that is expected for such a rapidly rotating M-dwarf.
The RV measurements, circular orbit fit and EW(H$\beta$) measurements are
shown in Fig.~\ref{J1042Fig}. The H$\beta$ emission line has a similar width
to the CCF and shows the same RV variation so it originates from the M-dwarf.
There is a clear cosine-like variation of EW(H$\beta$) with orbital phase, as
expected if it is due to the irradiation of one side of the M-dwarf by a hot
companion.  We applied a rotational broadening function to the spectra of
HD\,79210 (M0V) for various values of the projected equatorial rotational
velocity, \Vsini\ where $i$ is the inclination, and measured the FWHM of the
resulting CCF. From this calibration we estimate that the FWHM of 180\kms\
observed for the CCF of J1042+6442 corresponds to \Vsini\,$\approx 90\kms$. This
corresponds to a ``projected radius'' R$\sin i\approx 0.36\Rsolar$.  We have
compared the colour of this star to the colour-colour diagrams of
\citet{2008A&A...486..843A}. From the $(r-i)$ v. $(i-z)$ diagram (their
Fig.~2) we estimate  a spectral type of $\approx$M1 for J1042+6442. This
spectral type is consistent with the appearance of the spectrum shown in
Fig.~\ref{spectra}, e.g., the weak TiO bandhead near 4950\AA\ is slightly
stronger than the M0V star HD\,79210. The mass of an M1V star is about
0.5\Msolar\ and the radius is about 0.5\Rsolar. Combined with the mass
function $f_m = 0.093\Msolar$, we estimate a system inclination $i \approx
45^{\circ}$ so the mass for the unseen companion to the M1 star $\approx
0.75\Msolar$. There is large scatter in the relationship  between mass, radius
and spectral type for M-dwarfs, but these estimates are quite consistent with
the companion being a white dwarf. 

 The radius of the Roche lobe in this system is $\approx 0.5\Rsolar$, which is
comparable to the radius expected for an M1-type dwarf star. This raises the
possibility that there is accretion onto the companion from the M-dwarf
through the inner Lagrangian point.  However, there is no evidence for an
accretion disc or other accretion structures from the H$\beta$ emission line.
J1042+6442 has been detected in the ROSAT All Sky Survey
\citep{2000IAUC.7432R...1V} with a count rate of
$0.046\pm0.011\,\mathrm{cts/s}$ and with a hardness ratio
$\mathrm{HR1}=-0.29\pm0.20$, implying a relatively soft source. The
galactic column density in the direction of J1042+6442 is
$1.2\times10^{20}\,\mathrm{cm^{-2}}$ \citep{1990ARA&A..28..215D}. Given
the short distance, 67\,pc, we assume that about half of that column
is in front of J1042+6442. Adopting a thermal Bremsstrahlung spectrum
and a plasma temperature in the range $1-10$\,keV results in an X-ray
luminosity of J1042+6442 of
$L_\mathrm{X}=(2-6)\times10^{29}\,\mathrm{erg/s}$, which is roughly an
order of magnitude lower than the typical X-ray luminosity of
cataclysmic variables \citep{1997A&A...327..602V}, but is consistent with
the X-ray luminosity of rapidly rotating low-mass stars
\citep{2003A&A...397..147P}. More specifically, adopting a 0.5\,keV
Raymond-Smith plasma, appropriate for coronal emission, results in
$L_\mathrm{X}=2\times10^{29}\,\mathrm{erg/s}$, which is very similar to
the X-ray luminosities of the eclipsing short-period low-mass stars
GU\,Boo, YY\,Gem, and CU\,Cnc \citep{2003A&A...398..239R,
  2005ApJ...631.1120L}. We conclude that the X-ray emission from
J1042+6442 is consistent with coronal emission from the main-sequence
component, corroborating the detached 
pre-CV nature of the system.

\begin{table}
\caption{Radial velocity and EW(H$\beta$) measurements for J1042+6442
\label{J1042RV}}
\begin{center}
\begin{tabular}{@{}lrr}
\hline
\multicolumn{1}{l}{MJD}
&\multicolumn{1}{c}{$V_r$} & \multicolumn{1}{c}{$-$EW(H$\beta$)} \\
&\multicolumn{1}{c}{(\kms)} & \multicolumn{1}{c}{(\AA)} \\
\hline
  54823.2846   &$    112.0    \pm  3.7 $   &    4.38      $\pm$  0.27 \\ 
  54826.1419   &$   -163.7    \pm  2.9 $   &    3.78      $\pm$  0.24 \\ 
  54826.2602   &$     91.4    \pm  1.8 $   &    5.48      $\pm$  0.24 \\ 
  54826.2638   &$     78.3    \pm  3.0 $   &    5.52      $\pm$  0.25 \\ 
  54826.2887   &$    -64.9    \pm  2.6 $   &    5.29      $\pm$  0.26 \\ 
  54826.2923   &$    -93.3    \pm  2.8 $   &    4.80      $\pm$  0.25 \\ 
  54826.2959   &$   -119.4    \pm  2.4 $   &    5.59      $\pm$  0.27 \\ 
  54826.2995   &$    -91.5    \pm  2.9 $   &    5.50      $\pm$  0.27 \\ 
  54826.3032   &$   -117.0    \pm  2.6 $   &    5.55      $\pm$  0.29 \\ 
  54843.1881   &$     26.2    \pm  5.3 $   &    5.87      $\pm$  0.67 \\ 
  54843.2356   &$    142.9    \pm  8.2 $   &    7.48      $\pm$   1.0 \\ 
  54843.2643   &$     45.8    \pm 15.0 $   &                    ---   \\ 
  54843.2796   &$    -57.9    \pm  8.6 $   &    5.32      $\pm$   1.1 \\ 
  54843.9713   &$     15.6    \pm  3.6 $   &    2.09      $\pm$  0.36 \\ 
  54844.0278   &$    184.3    \pm  4.7 $   &    3.50      $\pm$  0.21 \\ 
  54844.1461   &$    -76.2    \pm  3.7 $   &    3.20      $\pm$  0.25 \\ 
  54844.2518   &$     76.6    \pm  5.0 $   &    4.84      $\pm$  0.58 \\ 
  54844.3147   &$   -147.1    \pm  6.6 $   &    1.16      $\pm$  0.72 \\ 
  54844.9890   &$    155.1    \pm  2.4 $   &    2.04      $\pm$  0.28 \\ 
  54845.0185   &$    145.3    \pm  2.5 $   &    4.01      $\pm$  0.21 \\ 
  54845.0760   &$    -79.2    \pm  2.8 $   &    4.84      $\pm$  0.32 \\ 
  54845.1065   &$   -166.5    \pm  4.8 $   &    3.26      $\pm$  0.20 \\ 
  54845.2039   &$    172.0    \pm  2.9 $   &    3.39      $\pm$  0.20 \\ 
  54845.2304   &$    118.9    \pm  2.7 $   &    3.76      $\pm$  0.20 \\ 
  54845.2375   &$     77.6    \pm  2.3 $   &    4.98      $\pm$  0.22 \\ 
  54845.2774   &$   -134.4    \pm  4.1 $   &    3.17      $\pm$  0.37 \\ 
\noalign{\smallskip}
\hline
\noalign{\smallskip}
\end{tabular}
\end{center}
\end{table}

\begin{table}
\caption{Circular orbit least-squares fit to our RV measurements of J1042+6442.
The model is $V_r = \gamma + K \sin[2\pi (T-T_0)/P]$ where $T$ is the time of
mid-exposure. The effect of orbital smearing due to the finite exposure time
have been included. The quantity $\sigma_{\rm ex}$ is added in quadrature to
the standard errors in Table~\ref{J1042RV} prior to calculating the
least-squares fit and is chosen so that $\chi^2\approx N_{\rm df}$, the number
of degrees of freedom in the fit.
\label{J1042Orbit}}
\begin{center}
\begin{tabular}{@{}lr}
\hline
\multicolumn{1}{@{}l}{Parameter} & \multicolumn{1}{l}{Value} \\
\hline
Period (days) & 0.197669 $\pm$ 0.000022 \\
T$_0$ (HJD)   & 2454834.7820  $\pm$ 0.0012 \\
$\gamma$ (\kms) & $ -0.1 \pm 4.4 $ \\
$K$ (\kms) & 166.7  $\pm$ 5.3 \\
$\sigma_{\rm ex}$ (\kms) & 18 \\
$N_{\rm df}$ & 22 \\
$\chi^2$ & 21.8 \\
\noalign{\smallskip}
\hline
\noalign{\smallskip}
\end{tabular}
\end{center}
\end{table}

\subsection{White dwarf effective temperatures \label{WDTeffSec}}
 We have estimated the  effective
temperature of the white dwarfs in the three new PCEBs we have identified
based on the available SDSS and GALEX photometry. We 
first use the SDSS $i-z$ colour of the system and the calibration of
\cite{2007AJ....134.2398C} to estimate the spectral type of
the dwarf star in these binaries. We then use the observed $z$-band magnitude
together with the $M_J$ and $z-J$ calibration from
\citeauthor{2007AJ....134.2398C} to estimate the distance to the dwarf star.
J-band magnitudes are taken from the 2MASS catalogue
\citep{2006AJ....131.1163S}. 
The contribution of the white dwarf at these wavelengths  is negligible.
Synthetic spectra for the white dwarf generated using the models described in
\citet{2005A&A...439..317K} were then  matched ``by-eye'' to the observed
GALEX FUV and NUV fluxes by varying the effective temperature assuming the
white dwarf has a surface gravity $\log g = 8.0$. The results are given in
Table~\ref{WDTeffTable}. It is difficult to make an accurate estimate of the
uncertainty in these estimates. For example, the three FUV fluxes reported for
J1042+6442 in GALEX GR4 are $139\pm  11$, $112.2  \pm 2.7$ and  $133.1 \pm
3.9$, which are clearly not consistent with a single value. We also found
differences of 10\,--\,20 percent when we compared the distances estimated
using the calibration of \citeauthor{2007AJ....134.2398C} to those derived
using the calibrations of \cite{2006PASP..118.1679D}. Changing the assumed
value of $\log g$ by $\pm 0.5$ changes the estimate of the effective
temperature by about $\pm 1000$K for J0941+4706 and J1042+6442, and $\pm
2000$K for J1016+4935. Given these factors, our opinion is that a fair
estimate of the errors in the effective temperatures of the white dwarfs is a
few thousand Kelvin.

\begin{table}
\caption{Estimate of the white dwarf effective temperature (T$_{\rm eff,WD}$)
based on the SDSS and GALEX for our three newly discovered PCEBs. The spectral
type (SpTy) and distance ($d$) of the dwarf star based on the observed $i-z$,
colours and $z$ magnitudes are given in columns 2 and 3, respectively. 
\label{WDTeffTable}}
\begin{center}
\begin{tabular}{@{}lrrrrrr}
\hline
\multicolumn{1}{@{}l}{Star} & \multicolumn{1}{l}{SpTy} &
\multicolumn{1}{@{}c}{$d$} & \multicolumn{1}{l}{ T$_{\rm eff,WD}$}&
\multicolumn{1}{@{}c}{$i$} & \multicolumn{1}{@{}c}{$i-z$} & 
\multicolumn{1}{@{}c}{$z-J$} \\
&&
\multicolumn{1}{@{}c}{(pc)} & 
\multicolumn{1}{@{}c}{(K)} \\
\hline
J0941+4706 & K7V  &  280  &   12000  & 14.05  & 0.31   & 1.19\\
J1016+4935 & M2V  &  190  &   14400  & 13.96  & 0.36   & 1.22\\
J1042+6442 & M3V  &   67  &    9800  & 12.26  & 0.54   & 1.33\\
\noalign{\smallskip}
\hline
\noalign{\smallskip}
\end{tabular}
\end{center}
\end{table} 

%

\section{Discussion}

 The survey strategy we have adopted has identified one new PCEB (J1042+6442)
and two stars that are also likely to be PCEBs (J0941+4706, J1016+4935) based
on the amplitude and timescale of the RV variability and the narrow width of
their H$\beta$ emission lines. These new identifications are in addition to
two CVs (PQ~Gem and BZ~UMa) and one pre-CV (BE~UMa) in the sample that satisfy
our selection criteria. Further observations will be required to determine the
basic parameters of J0941+4706 and J1016+4935. We have compared the colours of
these two stars to the colour-colour diagrams of \citet{2008A&A...486..843A}.
The dwarf star dominates the flux at these wavelengths so the colours are
indistinguishable from those of single stars.  The other colours of J0941+4706
are also indistinguishable from those of a single star according to the
criteria of \citeauthor{2008A&A...486..843A}, i.e., this is a PCEB that cannot
be identified from its SDSS colours alone. The colours of J1016+4935 and
J1042+6442 do fall within the selection criteria of
\citeauthor{2008A&A...486..843A} in the $(u-g)$ v. $(g-r)$ plane (although
they are near the limit), but not the $(g-r)$ v. $(r-i)$ plane. None of these
stars appear in the catalogue of \citeauthor{2008A&A...486..843A} because they
exceed the brightness limit of $g=16.5$ set by these authors to avoid stars
with unreliable photometry. The effective temperatures we have estimated for
the white dwarfs in these binaries (Section~\ref{WDTeffSec}) are much lower
than the temperature of the coolest WD that can be detected from a U-band
excess given by \cite{2003A&A...406..305S}, which supports our conclusion that
the combination of  GALEX and SDSS photometry can be used to identify PCEBs
that are difficult or impossible to identify using SDSS photometry alone.

 Rather than setting a brightness limit, we have used the flags provided with
the SDSS photometry to identify stars with reliable photometry. The brightness
at which saturation of the CCD will produce unreliable photometry  will depend
on the filter under consideration, the colour of the star and the seeing
during the observation. It is noticeable that the SDSS photometry for several
of our targets were obtained in worse-than-average seeing conditions. A full
exploration of how these factors affect our survey is beyond the scope of this
paper.  

 Two of the stars we have observed (J0754+2035 and  J0851+2903) are very
rapidly rotating K-dwarfs with strong chromospheric emission that show
small shifts in their measured RVs. These may be the result of wind-accretion
induced rapid rotation \citep[WIRRing stars; ][]{1996MNRAS.279..180J}. The
FUV-excess in these stars would then be due to the white dwarf remnant of the
asymptotic giant branch (AGB) star from whose wind the K-dwarf accreted both
material and angular momentum. The current separation of the WD and K-dwarf
can be large in this scenario ($\sim 100$au) so it may be possible to directly
observe these WD using high resolution imaging. The apparent RV variability of
these stars is likely to be dominated by spurious shifts associated with their
extreme magnetic activity, i.e., star-spots, and is unlikely to be due to
orbital motion. J0851+2903 can be identified as a binary star containing a WD
using the criteria of \citeauthor{2008A&A...486..843A} in the $(g-r)$ v.
$(r-i)$ plane, but J0754+2035 would be indistinguishable from a single K-dwarf
using the SDSS photometry alone.

The majority of the stars in our sample appear to be normal late-type dwarfs.
There is little contamination from unidentified quasars as a result of our
proper motion selection. The FUV excess in the targets that do not show radial
velocity shifts in our data may be due to a hot subdwarf of hot white dwarf
companion to the late-type star in wide orbit. High resolution imaging or RV
measurements with greater precision than those use here could be used to
confirm this hypothesis.

\section{Conclusions}
 
 We have shown that complementing SDSS photometry with GALEX photometry will
enable us to identify new PCEBs and WIRRing stars that cannot be identified
from SDSS photometry alone. The new binary stars we have identified with our
follow-up spectroscopy contain late-K or early-M dwarf stars. A more complete
survey of targets identified using the methods we have developed here will be
required to show whether or not the lack of PCEBs  containing dwarf stars
earlier than late-K is a real feature of this population.

\section*{Acknowledgments}
This research has made use of the SIMBAD database, operated at CDS,
Strasbourg, France. PFLM would like to acknowledge discussions with
R.D.~Jeffries that have clarified several issues with regard to WIRRing stars.
This publication makes use of data products from the Two Micron All Sky
Survey, which is a joint project of the University of Massachusetts and the
Infrared Processing and Analysis Center/California Institute of Technology,
funded by the National Aeronautics and Space Administration and the National
Science Foundation. We thank an anonymous referee for their careful reading of
the manuscript.

\label{lastpage}
\bibliographystyle{mn2e}  
\bibliography{mybib}

\end{document}